\documentclass[11pt, conference]{IEEEtran}

\usepackage{amsmath,amssymb,amsthm,graphicx,enumerate}
\usepackage{tikz}
\usetikzlibrary{arrows,positioning,fit}

\newcommand{\eps}{\varepsilon}

\newcommand{\Ebb}{\mathbb{E}}
\newcommand{\Pbb}{\mathbb{P}}

\newcommand{\Ccal}{\mathcal{C}}

\newcommand{\Scal}{\mathcal{S}}

\newcommand{\Xcal}{\mathcal{X}}
\newcommand{\Ycal}{\mathcal{Y}}

\newtheorem{thm}{Theorem}

\newtheorem{defn}{Definition}

\title{Hybrid Codes Needed for Coordination over the Point-to-Point Channel}
\author{
\authorblockN{Paul Cuff and Curt Schieler}
\authorblockA{Department of Electrical Engineering \\
Princeton University}
}

\begin{document}
\maketitle

\begin{abstract}
We consider a new fundamental question regarding the point-to-point memoryless channel.  The source-channel separation theorem indicates that random codebook construction for lossy source compression and channel coding can be independently constructed and paired to achieve optimal performance for coordinating a source sequence with a reconstruction sequence.  But what if we want the channel input to also be coordinated with the source and reconstruction?  Such situations arise in network communication problems, where the correlation inherent in the information sources can be used to correlate channel inputs.

Hybrid codes have been shown to be useful in a number of network communication problems.  In this work we highlight their advantages over purely digital codebook construction by applying them to the point-to-point setting, coordinating both the channel input and the reconstruction with the source.\footnote{This work was partially supported by the National Science Foundation (NSF) through the grant CCF-1116013.}
\end{abstract}

\section*{Introduction}
In point-to-point communication, one is usually concerned with minimizing the distortion between the source sequence $S^n$ and the reproduction sequence $\hat{S}^n$. In common parlance, we desire necessary and sufficient conditions such that $\Ebb[d(S^n,\hat{S}^n)]\leq D$. In the language of coordination \cite{cuff10}, we seek necessary and sufficient conditions such that the empirical distribution of $(S^n,\hat{S}^n)$ is close to a given $P_{S\hat{S}}$ in total variation. Both of these notions consider the source sequence and reproduction sequence, but fail to include the channel input and output sequences, $X^n$ and $Y^n$. We wish to understand the nature of optimal communication in scenarios where we might want $X^n$ or $Y^n$ to be correlated with (or, coordinate with) $S^n$ or $\hat{S}^n$.  To that end, we investigate which empirical distributions of $(S^n,X^n,Y^n,\hat{S}^n)$ can be reliably achieved if we design our encoder and decoder optimally.

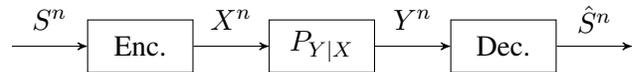
\begin{figure}
\begin{tikzpicture}
[node distance=1cm,minimum height=7mm,minimum width=14mm,arw/.style={->,>=stealth'}]
  \node[coordinate] (source) {};
  \node[rectangle,draw] (enc) [right =of source] {Enc.};
  \node[rectangle,draw] (ch) [right =of enc] {$P_{Y|X}$};
  \node[rectangle,draw] (dec) [right =of ch] {Dec.};
  \node[coordinate] (mhat) [right =of dec] {};

  \draw [arw] (source) to node[midway,above]{$S^n$} (enc);
  \draw [arw] (enc) to node[midway,above]{$X^n$} (ch);
  \draw [arw] (ch) to node[midway,above]{$Y^n$}(dec);
  \draw [arw] (dec) to node[midway,above]{$\hat{S}^n$} (mhat);
\end{tikzpicture}
\caption{{\em Point-to-point communication.}  An information source $S^n$ is encoded for communication through a noisy channel $P_{Y|X}$.  We consider how the channel input and output, $X^n$ and $Y^n$, can be correlated with the source and reconstruction, $S^n$ and $\hat{S}^n$.}
\label{figure setup}
\end{figure}

By understanding which empirical distributions can be achieved in the point-to-point communication problem, including the channel inputs and outputs, we gain insights that extend to network communication problems involving multiple correlated sources.  When information sources are correlated and communication occurs over channels with interference, it is useful to take advantage of the correlation and use it to strategically correlate the channel inputs to minimize the effect of the interference.  The point-to-point setting gives us a simple environment to explore the extent with which this is possible and the encoding and decoding techniques needed.

Despite the simplicity of the point-to-point setting, some interesting technology applications arise.  For example, a media broadcast, such as video, which is designed to accommodate both analog and digital receivers, fits well into this framework.  The most familiar approach in this case is to split the communication resources (bandwidth or power) into an analog transmission and a digital transmission, and to take advantage of the analog signal as side information for digital signal (as in the Wyner-Ziv setting).  However, it is not necessarily optimal to split the communication resources.  A single signal can serve both purposes, and in the context of coordination we want the channel output $Y^n$ and the digital reconstruction $\hat{S}^n$ to each satisfy different distortion constraints.  Another interesting application is the design of ``systematic'' transmissions, where an analog representation of the source information is designed to be robust to noise.  This fits into our framework -- we simply ask for coordination that results in the reconstruction $\hat{S}^n$ being equal to the channel input $X^n$.  Finally, digital watermarking can also be studied in this coordination framework.  In this case, the watermarked media is the channel input, which should obviously be highly correlated with the original media content.  Additionally, the watermark $(\hat{S}^n)$ should be detectable even if the media is altered in a limited way.

A classic result in information theory is the source-channel separation theorem, which says that if we have designed the optimal source codec and channel codec, then we can minimize average distortion by concatenating the two operations, using ``bits'' as glue. For our problem, however, separation is not good enough. We draw on recent results on hybrid codes to strictly improve the optimal regions. Hybrid codes combine bits, which are digital entities, with symbol-by-symbol operations, which are analog entities. These analog/digital combinations have been studied recently in the context of joint source-channel coding for discrete memoryless sources and channels and have been applied to multiuser scenarios in \cite{lim10} and \cite{minero11}.

Our problem is closely related to state amplification \cite{kim08}. There, $S^n$ is again given by nature, but the channel is given by $P_{Y|XS}$, not $P_{Y|X}$. The receiver's goal is to reproduce the state sequence, in contrast to the Gel'fand-Pinsker problem where the receiver wants to identify the sender's message. To use such a channel, $X^n$ must be coordinated with $S^n$; hence the relevance to our problem. The relation to state amplification and the Gel'fand-Pinsker problem prompts us to also consider the scenario where the encoder is constrained to be causal. That is, the input to the channel can only depend on past source symbols, as opposed to depending on the entire source sequence. In both this case and the strictly causal case, we identify the coordination region, using methods similar to those used in the analysis of causal state amplification in \cite{chouhuri10} and \cite{choudhuri11}.

\section*{Setup}

We are given an i.i.d. source parametrized by $P_S$ and a memoryless stationary channel parametrized by $P_{Y|X}$. The alphabets for source symbols, channel inputs, and channel outputs are denoted by $\Scal$, $\Xcal$, and $\Ycal$, respectively. In accordance with Figure~\ref{figure setup}, we have at our disposal an encoder and decoder; depending on the scenario, the encoder might be noncausal, causal, or strictly causal. For block length $n$, the encoder and decoder together constitute a code $(f^n,g^n)$ where $f^n$ and $g^n$ are sequences of functions $\{f_i\}_{i=1}^n$ and $\{g_i\}_{i=1}^n$:
\begin{itemize}
 \item noncausal encoder: $f_i:\Scal^n\rightarrow \Xcal$.
 \item causal encoder: $f_i:\Scal^i\rightarrow \Xcal$.
 \item strictly causal encoder: $f_i:\Scal^{i-1}\rightarrow \Xcal$.
 \item decoder: $g_i:{\cal Y}^n\rightarrow {\cal \hat{S}}$.
\end{itemize}
In the three cases above, at time $i$ the encoder produces the channel input $X_i$. After $n$ uses of the memoryless channel, the (noncausal) decoder produces $\hat{S}^n$ according to $\hat{S}_i=g_i(Y^n)$. The actions of the encoder and decoder, combined with the nature of the source and channel, give us an induced distribution on $(S^n,X^n,Y^n,\hat{S}^n)$.

Our model being defined, we can now state the condition for achievability.
\begin{defn}
A distribution $P_{SXY\hat{S}}$ is achievable if for all $\epsilon>0$ there exists a block length $n$ and a code $(f^n,g^n)$ such that
$$\mathbf{P} \; \left( \left\| P_{S^nX^nY^n\hat{S}^n}-P_{SXY\hat{S}}\right\|_{TV} > \epsilon \right) \; < \; \epsilon,$$
where $P_{S^nX^nY^n\hat{S}^n}$ is the empirical distribution of the tuple $(S^n,X^n,Y^n,\hat{S}^n)$ induced by the chosen code.
\end{defn}
It's not too hard to see that an achievable $P_{SXY\hat{S}}$ will necessarily factor like $P_SP_{X|S}P_{Y|X}P_{\hat{S}|SXY}$, where $P_S$ and $P_{Y|X}$ are the given source and channel parameters. This will be the case because $P_S$ and $P_{Y|X}$ are given by nature, and by the law of large numbers they will be close to the empirical distributions $P_{S^n}$ and $P_{Y^n|X^n}$ for sufficiently large $n$.

We can view proximity in total variation as akin to typicality.
\begin{defn}
 Given a distribution $P_{XY}$, the pair $(x^n,y^n)$ is in the $\eps$-typical set $T_\eps(X,Y)$ if
$$\lVert P_{x^n,y^n}-P_{XY}\rVert_{TV}\leq \eps$$
where
$$P_{x^n,y^n}(x,y)=\frac{1}{n}\sum_{i=1}^n 1\{x_i=x,y_i=y\}$$
is the empirical distribution of $(x^n,y^n)$.
\end{defn}
This is not too far from the usual definition of strong typicality. With this definition, we see that achieving $P_{SXY\hat{S}}$ is nothing more than ensuring that, $\forall \eps>0$, $(S^n,X^n,Y^n,\hat{S}^n)\in T_\eps(SXY\hat{S})$ with high probability. Next we define the coordination region, which will be the focus of our results.

\begin{defn}
The coordination region $\Ccal$ is the set of achievable $P_{SXY\hat{S}}$.
\end{defn}
It is not hard to prove that $\Ccal$ is a closed, convex set.

\section*{Results}
We begin with an inner bound on the coordination region for the case of noncausal encoder.
\begin{thm}[Noncausal encoder]
Let $P_S$ and $P_{Y|X}$ be the given source and channel parameters. When the encoder is allowed to be noncausal, an inner bound for the coordination region $\Ccal$ is given by
$$\left\{
\begin{array}{l}
 P_{SXY\hat{S}}:\vspace{.2cm}\\
 \exists U \text{ s.t. }\\
 P_{SXY\hat{S}U}=P_SP_{U|S}P_{X|US}P_{Y|X}P_{\hat{S}|UY},\\
 I(U;S)\leq I(U;Y).
 \end{array}
\right\} \subset \Ccal$$
Additionally, $P_{X|SU}$ and $P_{\hat{S}|UY}$ can be restricted to functions $x(s,u)$ and $\hat{s}(u,y)$ instead of general conditional distributions.
\end{thm}
This inner bound is a by-product of the achievability scheme for hybrid codes (see \cite{lim10},\cite{minero11}).  For more discussion on this, see the achievability section below.

The next two theorems give a complete characterization of the coordination region when the encoder is causal or strictly causal.
\begin{thm}[Causal encoder]
Let $P_S$ and $P_{Y|X}$ be the given source and channel parameters. With causal encoder, the coordination region $\Ccal$ is given by
$$\Ccal=\left\{
\begin{array}{l}
 P_{SXY\hat{S}}:\vspace{.2cm}\\
 \exists U,V \text{ s.t. }\\
 P_{SXY\hat{S}UV}=P_SP_{U}P_{V|SU}P_{X|SU}P_{Y|X}P_{\hat{S}|UVY},\\
 I(U,V;S)\leq I(U,V;Y).
 \end{array}
\right\}$$
Additionally, $P_{X|SU}$ and $P_{\hat{S}|UVY}$ can be restricted to functions $x(s,u)$ and $\hat{s}(u,v,y)$ instead of general conditional distributions.
\end{thm}

\begin{thm}[Strictly causal encoder]
Let $P_S$ and $P_{Y|X}$ be the given source and channel parameters. With strictly causal encoder, the coordination region $\Ccal$ is given by
$$\Ccal=\left\{
\begin{array}{l}
 P_{SXY\hat{S}}:\vspace{.2cm}\\
 \exists V \text{ s.t. }\\
 P_{SXY\hat{S}V}=P_SP_XP_{V|XS}P_{Y|X}P_{\hat{S}|VY},\\
 I(X,V;S)\leq I(X,V;Y).
 \end{array}
\right\}$$
Additionally, $P_{\hat{S}|VY}$ can be restricted to be a function $\hat{s}(v,y)$ instead of a general conditional distribution.
\end{thm}

\section*{Example}
Now we consider an example of what can be achieved using hybrid codes that is not achievable using Shannon's separation method.  When the encoder is noncausal, applying the separation method yields the following set of achievable distributions:
$$\left\{P_{SXY\hat{S}}:
\begin{array}{l}
 P_{SXY\hat{S}}=P_SP_{\hat{S}|S}P_{X}P_{Y|X}\\
 I(S;\hat{S})\leq I(X;Y)
 \end{array}
\right\} \subset \Ccal$$
In particular, achievable $P_{SXY\hat{S}}$ are required to factor as $P_{S\hat{S}}P_{XY}$. In contrast, Theorem 1 allows for greater range of correlation among $S$, $X$, $Y$, and $\hat{S}$; the following is an example of this.

Let $P_S$ be $\text{Bern}(p)$ and $P_{Y|X}$ be $\text{BSC}(\eps)$. Consider a distribution $P_{SXY\hat{S}}$ that is consistent with $P_S$ and $P_{Y|X}$, and also satisfies $\Pbb[S\neq\hat{S}]=\Pbb[S\neq X]=d$.
In other words, we want our reproduction $\hat{S}$ \emph{and} our channel input $X$ to be close to $S$ in hamming distortion. Notice that these conditions do not fully specify the distribution.  We will now see that if $\eps\leq d\leq p\leq \frac12$, Theorem 1 guarantees achievability of $P_{SXY\hat{S}}$. We make no claims here of necessity, just sufficiency.

To show the claim, choose $X$ such that $\Pbb[S\neq X]=d$ and $I(X;S)$ is minimized. From binary rate-distortion, we know this means $X\sim \text{Bern}(\frac{p-d}{1-2d})$. Choose $U=X$ and $\hat{S}(U,Y)=X$. With these choices, we see that
$$P_{SXY\hat{S}U}=P_SP_{U|S}P_{X|US}P_{Y|X}P_{\hat{S}|UY}$$
because $\Pbb[S\neq\hat{S}]=\Pbb[S\neq X]=d$. Now we just need to check that $I(U;S)\leq I(U;Y)$. We can view our situation as $S=X\oplus Z_1$ and $Y=X\oplus Z_2$, where $Z_1\sim \text{Bern}(d)$ and $Z_2\sim \text{Bern}(\eps)$ are independent of $X$. Then we have
\begin{eqnarray*}
I(U;S) & = & I(X;S) \\
& = & H(S)-H(Z_1) \\
& \leq & H(Y)-H(Z_2) \\
& = & I(X;Y) \\
& = & I(U;Y).
\end{eqnarray*}
where $H(Z_2)\leq H(Z_1)$ is due to $\eps\leq d$ and $H(S)\leq H(Y)$ is due to the fact that the entropy of a random variable increases when passed through a binary symmetric channel, in this case $\text{BSC}(d\star\eps)$.
\section*{Achievability}
\begin{figure}
\centering
\includegraphics[width=.4\textwidth]{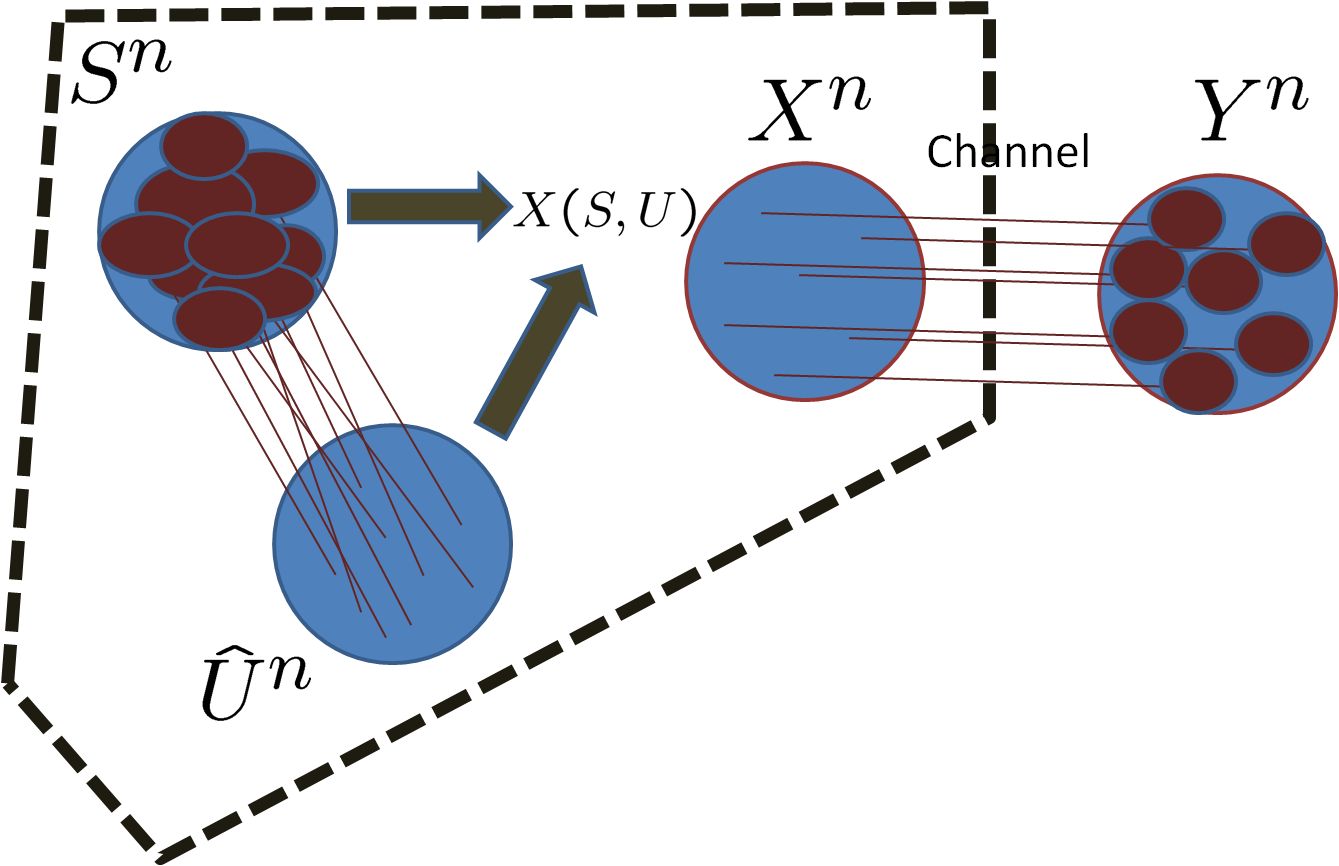}
\caption{{\em Hybrid code encoder.}  The encoder for a hybrid code has two steps in producing channel inputs form a source sequence $S^n$.  First, the source sequence is compressed using lossy compression and an auxiliary sequence $U^n$.  Then, the channel inputs $X^n$ are constructed symbol-by-symbol from $S^n$ and $U^n$.  If properly designed, the decoder is then able to decode $U^n$ from $Y^n$ and use both sequences to form the reconstruction $\hat{S}^n$.}
\label{figure achievability}
\end{figure}

The proof of achievability for the case of noncausal encoder, obtained by using hybrid codes, comes directly from \cite{lim10}, so we will omit it.  However, an illustration of hybrid codes is depicted in Figure~\ref{figure achievability}.  The main idea is to find an auxiliary $U^n$ sequence that is jointly typical with the source $S^n$ and is part of a sparse codebook that covers $\Scal^n$.  Then the channel input is constructed symbol-by-symbol as a function of $S$ and $U$.  This symbol-by-symbol function (the analog part) allows for $X^n$ to be correlated with $S^n$, while the construction of the auxiliary sequence (the digital part) takes advantage of compressing the entire source sequence.  The decoder obtains $U^n$ after observing $Y^n$ due to the sparsity of the codebook and then constructs $\hat{S}^n$ symbol-by-symbol from both $Y$ and $U$.

The achievability schemes for Theorems 2 and 3 are very similar to \cite{chouhuri10} and \cite{choudhuri11}, and we will now provide a sketch of the proof when the encoder is strictly causal (Theorem 3).

Fix $P_X$, $P_{V|XS}$ and $\hat{s}(v,y)$ consistent with an interior point of $\Ccal$. We will use block-Markov-like encoding, viewing the source as sequence of blocks:
$$\ldots,S^n(i-1),S^n(i),S^n(i+1),\ldots$$
The idea is to compress $S^n(i-1)$ in the $i$th block, using $X^n(i-1)$ and $Y^n(i-1)$ as side information. We first cover the source by using at least $2^{nI(S;V|X)}$ $V^n$ codewords and assign them randomly to greater than $2^{nI(S;V|X,Y)}$ bins (to first order in the exponent).  During the $i$th block we identify an index $\ell_i$ such that $V^n(\ell_i)$ is jointly typical with $S^n(i-1)$ and $X^n(i-1)$ and then transmit $X^n(m_i)$, where $m_i$ is the bin index of $V^n(\ell_i)$. If the number of bins does not exceed $2^{nI(X;Y)}$ and the size of each bin does not exceed $2^{nI(V;Y|X)}$, then the decoder will be able to recover $\ell_i$ by using $X^n(i-1)$ and $Y^n(i-1)$ as side information. Finally, the decoder produces $\hat{S}^n(i-1)$ symbol-by-symbol using $\hat{s}(v,y)$. Under the standard random coding analysis, this scheme ensures joint typicality with high probability as long as $I(X,V;S) < I(X,V;Y)$.

When the encoder is causal (but not strictly causal), the proof is similar, except now we introduce an additional auxiliary random variable $U$ that plays the role of $X$ in the strictly causal case. The channel input $X^n$ is produced symbol-by-symbol with a function $x(u,s)$.
\section*{Converse}
The converses for Theorems 2 and 3 are similar; we show the latter here. To prove the converse for Theorem 3, it is enough to show that for all $\eps>0$, there exists a distribution
$$\overline{P}_{SXY\hat{S}V}=P_S\overline{P}_X\overline{P}_{V|XS}P_{Y|X}\overline{P}_{\hat{S}|VY}$$
such that
\begin{itemize}
 \item $I(X,V;S)\leq I(X,V;Y)$
 \item $\lVert \overline{P}_{SXY\hat{S}}-P_{SXY\hat{S}}\rVert_{TV}<\eps$
\end{itemize}
Then by the closedness of $\Ccal$ we will be done. The key to the proof is in the identification of the auxiliary random variables. We first introduce $Q\perp(S^n,X^n,Y^n,\hat{S}^n)$ uniformly distributed on the set $\{1,...,n\}$, and define $V_Q=(S^{Q-1},Y_{Q+1}^n)$. Then we define $\overline{P}$ by setting $(S,X,Y,\hat{S})=(S_Q,X_Q,Y_Q,\hat{S}_Q)$ and $V=(V_Q,Q)$. With these choices, we can verify that $S\perp X$, $SV-X-Y$, and $SX-YV-\hat{S}$ all hold. Finally, we have
\begin{IEEEeqnarray*}{rCl}
 I(X,V;S) &=& I(X_Q,V_Q,Q;S_Q) \\
 &=& I(X_Q,V_Q;S_Q|Q) \\
 &=& \frac1n \sum_{i=1}^n I(X_i,V_i;S_i) \\
 &=& \frac1n \sum_{i=1}^n I(X_i,S^{i-1},Y_{i+1}^n;S_i) \\
 &=& \frac1n \sum_{i=1}^n I(S^{i-1},Y_{i+1}^n;S_i) \\
 &=& \frac1n \sum_{i=1}^n I(Y_{i+1}^n;S_i|S^{i-1}) \\
 &=& \frac1n \sum_{i=1}^n I(S^{i-1};Y_i|Y_{i+1}^n) \\
 &\leq& \frac1n \sum_{i=1}^n I(S^{i-1},Y_{i+1}^n;Y_i) \\
 &\leq& \frac1n \sum_{i=1}^n I(X_i,V_i;Y_i) \\
 &=& I(X_Q,V_Q;Y_Q|Q) \\
 &\leq& I(X_Q,V_Q,Q;Y_Q) \\
 &=& I(X,V;Y)
\end{IEEEeqnarray*}
To finish the proof, we can use our hypothesis that $P_{SXY\hat{S}}$ is achievable to show that
$$\left\| \overline{P}_{SXY\hat{S}}-P_{SXY\hat{S}} \right\|_{TV}<\eps.$$
The details of such a claim can be found in \cite{cuff10}, but the key is that
$$\Ebb[P_{S^n,X^n,Y^n,\hat{S}^n}]=P_{S_Q,X_Q,Y_Q,\hat{S}_Q}.$$

\section*{Summary}

Joint source-channel coding can be crucial, even in a point-to-point memoryless communication setting, when the right questions are asked.  The question we investigate is how strongly the channel input and output can be correlated with the information source during a communication transmission.  Our work makes use of the hybrid codes studied in \cite{lim10} and \cite{minero11} and finds that variations of hybrid analog-digital codes are optimal when the encoder is causal or strictly causal.

We find it interesting that we need not look further than the point-to-point memoryless communication setting to find use for unconventional and somewhat complex coding schemes such as hybrid codes.


\begin{thebibliography}{9}
 \bibitem{lim10} S.H. Lim, P. Minero, and Y.H. Kim, ``Lossy communication of correlated sources over multiple access channels,`` in \emph{Proc. 48th Allerton Conf. on Communication, Control, and Computing}, Oct. 2010.

 \bibitem{minero11} P. Minero, S.H. Lim, and Y.H. Kim, ``Joint source-channel coding via hybrid coding,'' in \emph{Proc. IEEE Int. Symp. on Info. Theory}, Saint Petersburg, Russia, Aug. 2011.

 \bibitem{kim08} Y.H. Kim, A. Sutivong, and T. Cover, ``State amplification,'' \emph{IEEE Trans. on Info. Theory}, vol. 54(5), pp.1850-1859, May 2008.

 \bibitem{chouhuri10} C. Choudhuri, Y.H. Kim, and U. Mitra, ``Capacity-distortion trade-off in channels with state,'' in \emph{Proc. 48th Allerton Conf. on Communication, Control, and Computation}, Oct. 2010.

\bibitem{choudhuri11} C. Choudhuri, Y.H. Kim, and U. Mitra. ``Causal state amplification,'' in \emph{Proc. IEEE International Symp. on Info. Theory}, Saint Petersburg, Russia, Aug. 2011.

\bibitem{cuff10} P. Cuff, H. Permuter, and T. Cover, ``Coordination capacity,'' \emph{IEEE Trans. on Info. Theory}, 56(9), pp.4181-4206, Sept. 2010.

\end{thebibliography}
\end{document}